\documentclass[11pt]{article}
\usepackage{amsmath, amssymb, graphicx, dsfont, xcolor, natbib, atbegshi, amsthm}
\usepackage[margin= 1in]{geometry}

\usepackage{amsmath, amssymb, atbegshi, dsfont, enumerate, graphicx, hyperref, natbib, prodint, setspace, xcolor, xr}

\newtheorem{theorem}{Theorem}[]

\newcommand{\ind}{\mathds{1}}
\newcommand{\Pe}{P_{\epsilon}}
\newcommand{\pe}{p_{\epsilon}}

\usepackage{graphicx}
\graphicspath{ {./images/} }
\usepackage{makecell}
\usepackage{float}
\usepackage{caption}

\begin{document}
\title{A class of nonparametric methods for evaluating the effect of continuous treatments on survival outcomes}

\author
{Yutong Jin$^{1}$, Peter B. Gilbert$^{1}$, Aaron Hudson$^{1}$ \\ $^{1}$ Fred Hutchinson Cancer Research Center}
\date{yjin2@fredhutch.org}

\maketitle


\begin{abstract}
In randomized trials and observational studies, it is often necessary to evaluate the extent to which an intervention affects a time-to-event outcome, which is only partially observed due to right censoring. For instance, in infectious disease studies, it is frequently of interest to characterize the relationship between risk of acquisition of infection with a pathogen and a biomarker previously measuring for an immune response against that pathogen induced by prior infection and/or vaccination. It is common to conduct inference within a counterfactual outcomes framework,  wherein we desire to make inferences about the counterfactual probability of survival through a given time point, at any given exposure level. To determine whether a causal effect is present, one can assess if this quantity differs by exposure level. Recent work shows that, under typical causal assumptions, summaries of the counterfactual survival distribution are identifiable both in randomized trials and observational studies. Moreover, when the treatment is multi-level, these summaries are also pathwise differentiable in a nonparametric probability model, making it possible to construct estimators thereof that are unbiased and approximately normal. This greatly facilitates inference. In cases where the treatment is continuous, the target estimand is no longer pathwise differentiable, rendering it difficult to construct well-behaved estimators without strong parametric assumptions. In this work, we extend beyond the traditional setting with multilevel interventions to develop approaches to nonparametric inference with a continuous exposure. We introduce methods for testing whether the counterfactual probability of survival time by a given time-point remains constant across the range of the continuous exposure levels. The performance of our proposed methods is evaluated via numerical studies, and we apply our method to data from a recent pair of efficacy trials of an HIV monoclonal antibody.

\end{abstract}







\label{firstpage}

\section{Introduction}
In observational studies and randomized trials, researchers often seek to establish a causal relationship between an exposure of interest and a time-to-event outcome.
This question can be framed as a statistical inference problem.
Using the counterfactual outcomes framework, we consider the counterfactual survival time, or the survival time that hypothetically would have occurred if an individual had been exposed to a specific exposure level.
One can assess whether there exists a causal relationship between the exposure and survival time by determining whether the counterfactual probability of survival prior to a given time point is the same across all exposure levels.
Our work discusses hypothesis testing approaches for equality of the counterfactual survival probabilities.
We consider the setting in which the outcome is potentially right-censored, and study participants can be lost to follow-up before event time.
We desire to make comparisons among the exposure-specific counterfactual survival probabilities under a nonparametric statistical model, to avoid risk of bias that can be induced by model misspecification.

There exist many methods for nonparametric inference on equality of the counterfactual survival probabilities when the exposure is multi-level.
In randomized controlled trials (RCTs) wherein the exposure is randomly assigned by the investigator, counterfactual survival probabilities for each exposure group can be estimated using the Kaplan-Meier estimator.
Kaplan-Meier estimators are unbiased and approximately normal, so it is straightforward to perform tests of equality of the counterfactual survival probabilities by comparing the exposure-specific survival estimates.

In observational studies, where the exposure is not under control of the investigator, inferring causal effects becomes more challenging due to the potential presence of confounding factors which affect both the exposure and the outcome. However, there exist covariate adjustment strategies that enable inference to be performed under standard causal assumptions.
One class of methods is based on estimation of the conditional distribution of the survival outcome given the covariates \citep{makuch1982adjusted}.
Inverse probability weighting is an alternative approach which uses estimators for the conditional distribution of the exposure given the covariates \citep{rosenbaum1983central,rosenbaum1987model}.
In order for unbiased and approximately normal estimators of the survival probability to be obtained, both strategies often rely on strong modeling assumptions \citep{cole2004adjusted,austin2014use}.
These strategies are therefore unsuitable for nonparametric testing.
For instance, it is often assumed that the conditional distribution of the outcome obeys a proportional hazards assumption or that the conditional distribution of the exposure can be described using a logistic linear model.
Recently, however, doubly robust nonparametric approaches have gained more attention. These approaches require estimating of both the conditional survival distribution and the conditional exposure distribution, but do not require parametric assumptions for either \citep{hubbard2000nonparametric,zhang2012contrasting}. Other more classical nonparametric estimation and testing strategies also exist \citep{jiang2011covariate,cai2020one,ozenne2020estimation}.

Our work focuses on the setting in which the exposure is continuous and can take infinitely many values. This is particularly relevant in practical applications, such as infectious disease studies, where investigators aim to understand the causal relationship between immunological biomarkers and risk of infection acquisition.
Though multi-level methods which rely on parametric assumptions on the conditional survival distribution or the conditional exposure distribution can be easily extended to the continuous setting, conducting nonparametric inference is more challenging in this context.
This is because the nonparametric testing strategies discussed above are only applicable when a large sample of study participants is observed at each exposure level, while in the case of continuous exposures, a very small number of samples is typically available at any given level. 
One approach to inference in the continuous setting is to group the exposure into a finite number of ordinal categories and to apply one of the multi-level procedures described above. While this may simplify the analysis, it raises some issues.
For instance, the arbitrary choice of cutoffs for the categorization can reduce statistical power \citep{cohen1983cost,maxwell1993bivariate,royston2006dichotomizing}. 


To address the previously discussed limitations of existing methods for inference on the relationship between continuous exposures and time-to-event outcomes, we propose a class of nonparametric tests of the null hypothesis that the counterfactual survival probabilities are the same at every exposure level. Our work builds upon recent advances from \citet{westling2021nonparametric} and \citet{hudson2023approach}, which developed frameworks for testing whether the counterfactual mean of an uncensored outcome takes the same value at every level of a continuous exposure. 
These methods are based on making inference on linear functionals of the dose-response function, which is defined as the map from an exposure level to its corresponding counterfactual mean. They show that it is possible to construct unbiased and asymptotically normal estimators for any linear functional of the dose-response function in a nonparametric model. They convert the original null into an equivalent statement that all appropriately centered linear functionals are zero, and then test this null by evaluating the maximum of all such linear functionals.
Our work extends this approach to the right-censored setting by leveraging theoretical results from \citet{westling2023inference}, which discusses efficient nonparametric estimation of the counterfactual survival function for a binary exposure.


\section{Preliminaries} 
\label{sec:notation}

\subsection{Data structure and notation}

For a positive integer $d$, let $W \in \mathcal{W} \subseteq \mathbb{R}^d$ be a $d$-dimensional vector of covariates, let $A \in \mathcal{A} \subset \mathbb{R}$ be a bounded exposure variable collected at baseline, let $ T >0$ denote the event time, and let $C \geq 0$ denote the censoring time. Additionally, let $T(a)$ represent the counterfactual event time under exposure level $a$, indicating the time to the event that would have been observed had an individual been exposed to $a$. For any $a$, $T(a)$ is assumed to be positive. Similarly, let $C(a)$ represent a counterfactual right-censoring time under exposure $a$. While we assume $T(a)$ is strictly positive, we allow for $C(a)$ to be zero or positive.
The complete data take the form $X = (W, A, T, C)\sim Q_0$, and we assume $n$ \textit{i.i.d.} observations $X_1, X_2, \ldots, X_n$ are generated from $Q_0$. In practice, we do not observe the complete data $X$. We instead observe the incomplete data $O_1, O_2, \ldots, O_n$, where $O = (W, A, Y, \Delta) \sim P$, $Y := \min\{T, C\}$, and $\Delta = \mathds{1}(T \leq C)$. We assume that $P$ belongs to a nonparametric model $\mathcal{M}$, which is unrestricted, aside from mild regularity conditions. Throughout, we use $E_{\tilde{P}}$ to denote the expectation under any distribution $\tilde{P} \in \mathcal{M}$.
In this study, we are interested in using the incomplete data to make inferences about the counterfactual probability of survival through a specified time point $t > 0$ under a given exposure $a$, denoted by $P_{Q_0}(T(a) > t)$. 
This quantity can be interpreted as the probability that an individual would experience an event beyond time point $t$ if hypothetically exposed to $a$. 

\subsection{Causal identification of the counterfactual survival probability} 
\label{sec:identification}
The counterfactual survival probability depends on the counterfactual survival time $T(a)$, which is not observable because (1) outcomes are right-censored and (2) every participant is only exposed to a single level $A$. 
Nonetheless, the counterfactual survival probability can be identified as a summary of the observed data distribution $P$ under a set of standard assumptions.
Let $g_P(a \mid w)$ denote the conditional density of $A$ given $W$ under $P$. 
We assume there exists $\tau > t$ such that the following conditions hold:
\begin{enumerate}
    \item[(A1)] \emph{Consistency}: $A=a$ implies $T=T(a)$ and $C=C(a)$.
    \item[(A2)] \emph{Ignorability}: (A2.1) $T(a)\ind[T(a) \leq \tau]$ is conditionally independent of $A$, given $W$; (A2.2) $C(a)\ind[C(a) \leq \tau]$ is conditionally independent of $A$, given $W$; (A2.3) $T(a)\ind[T(a) \leq \tau]$ is conditionally independent of $C(a)\ind[C(a) \leq \tau]$, given $A$ and $W$.
    \item[(A3)] \emph{Positivity}: There exist two constants $c_1$ and $c_2$, such that (A3.1) $g_P(A=a \mid w) > c_1$ for all $a\in \mathcal{A}$ and $w \in \mathcal{W}$; (A3.2) $c_2 < P(C(a) \geq \tau \mid W) < 1$ almost surely $P$.
    \item[(A4)] No interference between individuals, and no hidden variations of exposures.
\end{enumerate}
Assumption (A1) stipulates that the observed time to event or censoring for each observation corresponds to the counterfactual value under the observed exposure.
Assumptions (A2.1) and (A2.2) stipulate that conditional on $W$, any counterfactual event or censoring occurring prior to time $\tau$ is unrelated to the observed exposure $A$. These two conditions generally hold in randomized trials, where the exposure is randomly assigned, for any set of covariates $W$ (including an empty set of covariates). 
However, in observational studies, these conditional independence assumptions may not hold unless the covariates are selected carefully. In practice, these assumptions are unverifiable in observational studies.
Assumption (A2.3) stipulates that the event times and censoring times occurring through time $\tau$ are conditionally independent, given the exposure and covariates. Assumption 3 places two types of positivity requirements. First, (A3.1) assumes that the conditional density of the exposure given the covariates is positive for any set of exposure and covariate values. Second, (A3.2) ensures that there is a positive probability of uncensored observations occurring at or beyond time $\tau$ for each possible covariate profile. 

In what follows, we state a result that identifies the counterfactual survival probability as a functional of the observed data distribution $P$ \citep[Theorem 1 of][]{westling2023inference}. 
Let $\prodi$ denote the Riemann-Stieltjes product integral, and let $S_P(t \mid a,w) := \prodi_{(0,t]} \{ 1- \Lambda_P(du \mid a,w)\}$, 
$\Lambda_P(du \mid a,w) := \int_0^t F_{P,1}(du \mid a,w) / R_P(u \mid a,w)$, 
$F_{P,1}(t \mid a,w) := P(Y \leq t, \Delta=1 \mid A=a, W=w)$, and 
$R_P(t \mid a,w) := P(Y \geq t \mid A=a, W=w)$.
\begin{theorem}
For a given $a \in \mathcal{A}$, and $0 < t < \tau$, suppose that conditions A1-A4 hold. Then the counterfactual survival probability $P_{Q_0}(T(a) > t)$ is equal to
\[
\theta_{P}^a(t) := E_P \left[ S_P( t \mid a, W ) \right]  \ .
\] 
\end{theorem}
This theorem considers the identification of a smoothed survival estimator and provides a construction of counterfactuals by marginalizing over the observed data distribution, using the time-to-event regression model, via the g-computation formula \citep{gill2001causal}.
Any violation of Assumptions (A1)-(A4) may preclude a causal interpretation of the estimand $\theta_P^a(t)$. Even in cases where these assumptions are unmet, $\theta_P^a(t)$ maintains an interpretation as an expected conditional survival time, averaged over the covariate distribution. Therefore, $\theta_P^a(t)$ remains useful for measuring the conditional association between the exposure and the outcome, given the covariates.

\section{Problem statement and summary of proposed methodology}
\label{sec:overview}
Our objective is to determine whether the counterfactual survival probability, $\theta_P^a(t)$ takes the same value at every level of the exposure, for a single fixed $t$. This is equivalent to $\theta_P^a(t)$ being equal to its mean $E_P[\theta^A_P(t)]$ for all $a$.
Letting $\bar{\theta}_P^a(t) := \theta_P^a(t) - E_P[\theta_P^A(t)]$, we develop a test of the null,
\begin{align}
    H_0: \bar\theta_{P}^{a}(t) = 0, \hspace{1em} \text{for all } a \in \mathcal{A} \ .
    \label{eqn:flat_null}
\end{align}
In this section, we describe existing challenges to testing the null in \eqref{eqn:flat_null} and provide an overview of our proposed general test. 
We begin by discussing an approach to inference for cases where $A$ is discrete.
We then discuss difficulties with extending this approach to the continuous setting. This discussion leads us to consider a useful reformulation of the problem, which forms the basis of our general testing procedure.

First, suppose that $A$ is discrete and takes values in the set $\{1, 2, \ldots, K\}$. In this case, it is possible to construct a nonparametric estimator of $\theta_{P}^a(t)$ that has standard deviation decreasing at the rate of $n^{1/2}$, has bias approaching zero at a faster rate than $n^{1/2}$, and attains a normal limiting distribution. This is because $\theta_{P}^a(t)$ is pathwise differentiable, meaning it changes smoothly with small shifts from $P$ toward another probability distribution \citep{bickel1998efficient}. There exists a general theory for constructing efficient nonparametric estimators of pathwise differentiable estimands. In particular, estimation of counterfactual survival probabilities in the discrete exposure setting is carefully studied in \citet{westling2023inference}. 

Since the null holds only when $\bar{\theta}_{P}^a(t) = \theta_P^a(t) - E_P[\theta^A_P(t)] = 0$ for all $a$, one can construct a test by estimating any norm of $\{\bar{\theta}_{P}^1(t), \ldots, \bar{\theta}_{P}^K(t)\}$. Pathwise differentiability of $\theta_{P}^a(t)$ allows for construction of asymptotically normally estimators $\bar{\theta}_{n}^a(t)$ of $\bar{\theta}_{P}^a(t)$.
This in turn allows for construction of tests of $H_0$ based on, e.g., $\ell_1$ or $\ell_2$ norms of the estimators: 
\begin{align*}
    &T_{n,1} = \sum_{a \in \{1, 2, \ldots K\}} \left| \bar{\theta}_n^a(t)\right|, \qquad T_{n,2} = \left[\sum_{a \in \{1, 2, \ldots K\}} \left\{ \bar{\theta}_n^a(t)\right\}^2 \right]^{1/2}.
\end{align*}
The null limiting distribution of either test statistic can be characterized as the norm of a mean zero Gaussian random vector, making it straightforward to perform a hypothesis test.

The nonparametric testing approach above does not readily extend to the setting in which $A$ is continuous because $\theta_{P}^a(t)$ is no longer pathwise differentiable in the nonparametric model.
To provide some intuition, in the discrete setting, $\theta_P^a(t)$ is pathwise differentiable because it is an average of conditional survival probabilities, taken over a large sub-population with $A = a$. Observations with $A \neq a$ do not contribute unbiased information for estimating $\theta_P^a(t)$ in nonparametric models as we do not assume that there is a strong structural relationship between the exposure level and the counterfactual survival probability (e.g., we do not assume $a \mapsto$ $\theta_P^a(t)$ is log-linear).
However, $A = a$ with positive probability in the discrete setting, so the number of observations with $A = a$ tends to infinity as $n$ grows.
Hence, for sufficiently large $n$, one will have available a large sample with $A = a$.
Conversely, in the continuous setting where the probability of observing $A = a$ is infinitesimally small, one will not have access to a sufficiently large sample to construct an unbiased estimate of $\theta_P^a(t)$.

To address the challenges to inference caused by non-pathwise differentiability, we propose an approach that adapts the general framework for inference on non-pathwise differentiable estimands from \citet{hudson2021inference}. In particular, \citet{hudson2021inference} shows that it is possible to perform inference on function-valued estimands (in our case, the map $a \mapsto \theta_P^a(t)$) by constructing hypothesis tests based on estimation of linear contrasts of the parameter of interest, rather than its evaluation at any fixed exposure level. This idea has been applied to perform inference on the causal dose-response function in the continuous exposure setting with uncensored outcomes in \cite{westling2021nonparametric} and \citet{hudson2023approach}.
To make this explicit, for a bounded function $h$ from $\mathcal{A}$ to $\mathbb{R}$, we define a linear contrast
\[
\psi_{P,t} (h) = E_P \left\{ \bar{\theta}_P^A(t) \; h(A) \right\} \ .
\]
We first observe that because $\bar{\theta}^a_P(t)$ is zero under the null, $\psi_{P,t}(h)$ is necessarily zero for any $h$.
Conversely, if $\bar{\theta}_P^a(t)$ is non-zero on a subset $\mathcal{A}^*$ of $\mathcal{A}$ that has positive probability mass, there must exist an $h^*$ such that $ \psi_{P,t}(h^*)$ is non-zero.
Therefore, the null in \eqref{eqn:flat_null} can be reformulated as
\begin{align}
H_0: \hspace{2em} \psi_{P,t} (h) = 0 ,\hspace{1em} \text{for all bounded } h \ . \label{eqn:non-relaxed-null}
\end{align}
This reformulation of the null based on the linear contrasts $\psi_{P,t}(h)$ is useful because, as we will show in Section \ref{sec:estimation}, $\psi_{P,t}(h)$ is pathwise differentiable in the continuous exposure setting though $\theta^a_P(t)$ is not.
The contrast parameter $\psi_{P,t}(h)$ amalgamates information about the counterfactual survival for all exposure levels into a single estimand by evaluating a weighted average of $\theta_{P}^a(t)$ using all $a \in \mathcal{A}$. This allows all observations to contribute to unbiased estimation of $\psi_{P,t}(h)$.

In view of the above construction, a promising strategy for inference could consist of estimating $\psi_{P,t}(h)$ for many $h$ and determining whether there exists an $h^*$ for which $\psi_{P,t}(h^*) \neq 0$. Uniform estimation of $\psi(h)$ over the class of all bounded functions $h$ is challenging. However, it is feasible to uniformly estimate $\psi_{P,t}(h)$ within a class $\mathcal{H}$ that obeys mild structural constraints. We propose to test the relaxed null
\begin{equation}\label{eq:relaxed-null}
    \bar{H}_0: \hspace{2em} \Psi_{P,t}(\mathcal{H}) := \sup_{h \in \mathcal{H}}\left| \psi_{P,t} (h)\right| = 0.
\end{equation}
We show in Section \ref{sec:estimation} that when $\mathcal{H}$ is not overly complex, one can construct an estimator $\Psi_{n,t} (\mathcal{H})$, such that the supremum statistic $\sup_{h \in \mathcal{H}} n^{1/2} |\psi_{n,t}(h)|$ converges weakly to the supremum of a mean zero Gaussian process under the null hypothesis. This characterization of the null limiting distribution enables us to perform a hypothesis test with asymptotic type-1 error control at the nominal level.

The relaxed null $\bar{H}_0$ in \eqref{eq:relaxed-null} holds whenever the non-relaxed null $H_0$ in \eqref{eqn:non-relaxed-null} holds. However the converse is not necessarily true; $\bar{H}_0$ may still hold when $H_0$ does not. 
In other words, there may exist $h^*$ for which $\psi_{P,t}(h^*) \neq 0$, but $\mathcal{H}$ might be too small to contain such $h^*$.
This motivates us to consider classes $\mathcal{H}$ that are large, and possibly infinite-dimensional, so that we have power to detect a wide range of alternative hypotheses. However, we cannot allow the class to be so large as to render uniform estimation of $\psi_{P,t}(h)$ impossible. 
In Section \ref{sec:construction-of-H}, we will describe some possible constructions of $\mathcal{H}$ that can result in well-powered tests in realistic scenarios. 
We will see that the supremum estimand $\Psi_{P,t}(\mathcal{H})$ in \eqref{eq:relaxed-null} can be interpreted as a norm of $\bar\theta^a_{P}(t)$ for some choices of $\mathcal{H}$ under structural assumptions on $\theta^a_P(t)$. We emphasize that structural assumptions on $\theta^a_P(t)$ are not required for type-1 error control. These assumptions are only used to provide insight into what constructions of $\mathcal{H}$ can lead to well-powered tests when certain alternative hypotheses hold.
\section{Inference procedure}
\label{sec:estimation}
In this section, we discuss technical theoretical aspects of our proposed inference procedure. First, we discuss pathwise differentiability of $\psi_{P,t}(h)$ for fixed $h$, describe an efficient and uniformly consistent estimator for $\{\psi_{P,t}(h): h \in \mathcal{H}\}$ for $\mathcal{H}$ satisfying regularity conditions, and 
describe a hypothesis test based on estimation of $\Psi_{P,t}(\mathcal{H})$ in \eqref{eq:relaxed-null} for general $\mathcal{H}$. We then discuss how to construct $\mathcal{H}$ and interpret $\Psi_{P,t}(\mathcal{H})$ under different selections of $\mathcal{H}$. 
Proofs of all theorems are provided in the supplementary materials.

\subsection{\texorpdfstring{Estimation of $\psi_{P, t} (h)$}{Estimation of psi P,t (h)}}
We begin this section by describing the notion of pathwise differentiability more formally. 
Let $\tilde{P} \in \mathcal{M}$ be a probability distribution with density $\tilde{p}$, and let $\omega$ be a function of the observed data with mean zero under $\tilde{P}$.
Let $\tilde{P}_{\epsilon}$ be a parametric sub-model with density $\tilde{p}_\epsilon$ such that $\tilde{P}_{\epsilon} = \tilde{P}$ at $\epsilon = 0$, and its score function is $\frac{d}{d\epsilon} \log p_{\epsilon} = \omega$. 
An estimand $F: \mathcal{M} \to \mathbb{R}$ is pathwise differentiable at $\tilde{P}$ if there exists $\phi_{\tilde{P}}$ that does not depend on $\omega$ for which 
$\lim_{\epsilon\rightarrow0}\epsilon^{-1} \left\{ F\left(\tilde{P}_\epsilon\right) - F\left(\tilde{P}\right)\right\} = E_{\tilde{P}}[\phi_{\tilde{P}}(O)\omega(O)]$. 
We refer to $E_{\tilde{P}}[\phi(O)\omega(O)]$ as the pathwise derivative of functional $F$ at $\tilde{P}$ and refer to $\phi_{\tilde{P}}$ as the nonparametric efficient influence function of $F$. That a functional $F$ is pathwise differentiable essentially means that $F(\tilde{P})$ changes smoothly as we make small shifts away from $\tilde{P}$ along any sub-model that passes through $\tilde{P}$.
Characterizing the nonparametric efficient influence of a pathwise differentiable estimand is key for constructing efficient estimators and studying their asymptotic properties \citep{bickel1998efficient,hines2022demystifying}.

We next provide a result that establishes pathwise differentiability of $\psi_{P,t}(h)$. We first introduce some additional notation. 
As a counterpart to $S_P(t \mid a,w) = P(T >t \mid A=a,W=w)$, we let $G_P(t \mid a,w) := P(C \geq t \mid A=a, W=w)$ denote conditional probability  of censoring at time $t$ or later. We note that $S_P(t \mid a,w)$ is a right-continuous function of $t$, whereas $G_P(t \mid a,w,)$ is left-continuous. 
Recall from Section \ref{sec:identification} that $g_P$ and $\Lambda_P$ denote the conditional exposure density and the conditional cumulative hazard, respectively.
The following theorem states that $\psi_{P, t}(h)$ is pathwise differentiable in a nonparametric model and provides the explicit form of the nonparametric efficient influence function.
\begin{theorem}
For any fixed $h$, $\psi_{P,t} (h)$ is pathwise differentiable in a nonparametric model, and its efficient influence function under a model $P \in \mathcal{M}$ that satisfies (A1)-(A4) is
\begin{align*}
&D_{P}(w,a,y,\delta; h) \\
=& \left\{h(a) - E_P[h(A)]\right\} \bar{\theta}_{P}^{a}(t)
+ E_{P}\left[ S_{P}(t \mid A,w) \{h(A) - E_P[h(A)]\} \right] 
- 2E_{P}\{ h(A) \bar{\theta}_{P}^A(t) \} \\
+& \left\{h(a) - E_P[h(A)]\right\} \left\{ S_{P}(t \mid a,w) \frac{E_P(g_P(a \mid W))}{g_P(a \mid w)} \left[ H_P(t \wedge y, a, w) - \frac{\ind(y \leq t, \delta=1) S_{P}(y^- \mid a,w)}{S_{P}(y \mid a,w) R_{P}(y \mid a,w)} \right] \right\} \ ,
\end{align*}
\noindent where $ H_P(t \wedge y, a, w) = \int^{t\wedge y} \frac{\Lambda_P(du \mid a,w)}{S_P(u \mid a,w) \, G_P(u^- \mid a,w)}$ \ .
\end{theorem}

We are now ready to discuss the construction of estimators for $\psi_{P,t}(h)$. This requires the estimation of three nuisance parameters upon which $\psi_{P,t}(h)$ and its efficient influence function depend: $S_P$, $G_P$ and $g_P$. 
These nuisance parameters are not pathwise differentiable in a nonparametric model.
Consequently, it is not possible to obtain nonparametric estimators that are unbiased and converge at the parametric rate of $n^{1/2}$. 
Nonetheless, it is possible to obtain consistent estimators with a slower convergence by using flexible data-adaptive regression methods (e.g., machine learning).
These flexible nuisance estimators typically achieve consistency under weak structural assumptions on the true nuisance (e.g., smoothness) by balancing a bias-variance trade-off. Hence nuisance estimators may retain non-negligible asymptotic bias. 
Our theoretical framework accommodates the use of such flexible nuisance estimators, provided they satisfy certain regularity conditions, which we discuss later in this section. 
Specific nuisance estimators we consider are described in Section \ref{sec:implementation}. Hereafter, we assume estimators $S_n$, $G_n$ and $g_n$ are available.

We begin by describing a na\"ive plug-in estimator of $\psi_{P,t} (h)$, 
\begin{equation}\label{eq:plug-in psi}
\psi_{n,t} (h) = \frac{1}{n} \sum_{i=1}^n \left[ \left(\theta_{n}^{A_i}(t) - \frac{1}{n}\sum_{m=1}^n \theta_n^{A_m}(t)\right) \right] h(A_i) \ ,
\end{equation}
where $\theta_n^a(t) = n^{-1} \sum_{i=1}^n S_n(t \mid a, W_i)$ is the plug-in estimator of the counterfactual survival probability.
Bias from estimation of the conditional survival probability can propagate, leading to the plug-in estimator retaining non-negligible bias as well. This bias retention makes the plug-in estimator unsuitable for performing statistical inference. 

Fortunately, there exist strategies for correcting biased plug-in estimators. We apply the one-step bias correction method proposed by \citet{pfanzagl1982contributions}. This approach reduces the bias by adding an empirical average of the estimated efficient influence function to the initial plug-in estimator. 
Let $H_n(t \wedge Y_i,A_i,W_i)$ and $Z_n(O_i;h)$ be defined as
\begin{align*}
    &H_n(t \wedge Y_i,A_i,W_i) = \int^{t\wedge Y_i} \frac{\Lambda_n(du| A_i,W_i)}{S_n(u|A_i,W_i) \, G_n(u^- |A_i,W_i)} \ ,\\
    &Z_n(O_i;h) = \left\{h(A_i) -\frac{1}{n}\sum_{m=1}^n[h(A_m)]\right\}  \left\{ S_{n}(t|A_i, W_i) \frac{\frac{1}{n} \sum_{j=1}^n g_n(A_i|W_j)}{g_n(A_i|W_i)} \right\}.
\end{align*}
We define $D_n$ as a plug-in estimator for the efficient influence function, 
\begin{align}
    \label{eq:empirical-eif}
    D_n(O_i;h) = \left[H_n(t \wedge Y_i,A_i,W_i) - \frac{\ind(Y_i \leq t, \Delta_i=1) S_{n}(Y_i^- |A_i, W_i)}{S_{n}(Y_i|A_i, W_i) R_{n}(Y_i|A_i, W_i)} \right] Z_n(O_i;h) \ .
\end{align}
Finally, the one-step estimator is
\begin{align*}
    \psi_{n,t}^{\dagger} (h) 
    =& \psi_{n,t} (h) + \frac{1}{n}\sum_{i=1}^n D_{n}(O_i; h) \ .
\end{align*}

Under mild conditions, the one-step estimator is $n^{1/2}$-rate convergent and asymptotically normal, both at any fixed $h$ and uniformly over a class $\mathcal{H}$.
Prior to providing a formal result describing the one-step estimator's asymptotic properties, we discuss the working assumptions.
Suppose the following hold:
\begin{enumerate}
    \item[(B1)] There exists a $P$-Donsker class that contains $D_P(\cdot;h)$ and $D_n(\cdot;h)$ for each $h \in \mathcal{H}$ with probability tending to one.
    \item[(B2)] It holds that $\sup_{h\in \mathcal{H}} \int \left\{ \bar{\theta}_n^a(t) - \bar{\theta}_P^a(t) \right\}^2 dP = o_P(1)$.
    \item[(B3)] It holds that 
    $\int_0^t \left( \frac{g_P G_P(y| a,w)}{g_n G_n(y^-| a,w)} -1 \right) \left( \frac{S_P(y| a,w)}{S_n(y| a,w)} -1 \right) (dy| a,w) = o_P(n^{-1/2})$.
\end{enumerate}
Condition (B1) restricts the complexity of the collection of candidate estimators for the nuisance parameters and the function class $\mathcal{H}$. Many function classes, such as classes of functions with bounded variation norm, are known to satisfy the Donsker property. In general, Donsker assumptions can be verified by studying the metric entropy of the function class \citep{van2000asymptotic}. Conditions (B2) and (B3) state that the nuisance estimators must be convergent, though the convergence rates can be slower than $n^{1/2}$. 
In view of (B3), the estimator we propose enjoys a doubly-robust property: it remains efficient if $S_n$ is estimated at a relatively slow rate while $G_n$ and $g_n$ are estimated at a fast enough rate to compensate, or vice versa.
The theorem below establishes uniform weak convergence of the one-step estimator.
\begin{theorem} \label{thm:weak-convergence}
    Assuming (B1)-(B3), the one-step estimator $\psi_{n,t}^{\dagger} (h)$ has the representation
    \[\psi_{n, t}^{\dagger} (h) - \psi_{P,t} (h) = \frac{1}{n} \sum_{i=1}^n D_{P}(O_i; h) + r_n(h) \ ,\]
    with $\sup_{h\in \mathcal{H}} \left| r_n(h) \right| = op(n^{-1/2})$. Additionally, $\{[\psi_{n, t}^{\dagger}(h) - \psi_{P,t}(h)]: h \in \mathcal{H} \}$ converges weakly to a mean zero Gaussian process $\mathbb{G}$ with covariance $\Sigma_P: (h_1, h_2) \mapsto E[D_{P}(O;h_1) D_{P}(O;h_2)]$.
\end{theorem}
An important implication of Theorem \ref{thm:weak-convergence} is that under the null, when $\psi_{P,t}(h)$ is zero for all $h$, $\sup_{h \in \mathcal{H}} n^{1/2} \psi_n^{\dagger}(h)$ converges weakly to $\sup_{h \in \mathcal{H}} |\mathbb{G}(h)|$. 
We therefore have available an estimator for $\Psi_{P,t}(\mathcal{H})$ in \eqref{eq:relaxed-null} with a fully-characterized null limiting distribution. 
As stated in Section \ref{sec:overview}, this allows for straightforward hypothesis test construction.

To be explicit, we propose to test the relaxed null hypothesis $\bar{H}$ in \eqref{eq:relaxed-null} using the test statistic $\Psi^\dagger_{n,t}(\mathcal{H}) := \sup_{h \in \mathcal{H}}|\psi_{n,t}^\dagger(h)|$.
For $\alpha \in (0,1)$, let $\nu^*_{1-\alpha}$ denote the $1-\alpha$ quantile of the limiting distribution of $\sup_{h \in \mathcal{H}}|\mathbb{G}(h)|$.
A test that rejects the null when $n^{1/2}\Psi^\dagger_{n,t}$ exceeds $\nu^*_{1-\alpha}$ achieves type-1 error rate equal to $\alpha$ asymptotically.
For a realization $\nu$ of $n^{1/2}\Psi^\dagger_{n,t}(\mathcal{H})$, an asymptotic approximation for the p-value is $\rho(\nu) := P(\sup_{h \in \mathcal{H}}|\mathbb{G}(h)| > \nu)$.
The limiting distribution of the test statistic may not be available in closed form but can be approximated using Monte Carlo sampling.
We discuss implementation details in Section \ref{sec:implementation}.



\subsection{\texorpdfstring{Construction of $\mathcal{H}$}{Construction of H}
\label{sec:construction-of-H}}
The choice of $\mathcal{H}$ affects the test's type-1 error rate and power. Using a flexible class can result in a test that is well-powered against a wide range of alternatives. However, when $\mathcal{H}$ is too complex, type-1 error control may be compromised as uniform estimation of $\psi_{P,t}(h)$ may not be feasible.
To obtain a well-calibrated test, we construct $\mathcal{H}$ as a class of functions that satisfies two types of constraints.

First, we introduce a \emph{scale constraint}. 
We select some norm $\|\cdot\|$ on the space of functions from $\mathcal{A}$ to $\mathbb{R}$ and constrain all $h \in \mathcal{H}$ to satisfy $\|h\| \leq 1$. This ensures that the supremum in the null hypothesis \eqref{eq:relaxed-null} exists. Moreover, this constraint can enable $\Psi_{P,t}(\mathcal{H})$ to be interpreted as a norm of $\bar{\theta}_P^a(t)$. We consider two scale constraints as examples. 
First, suppose each $h$ has a bounded supremum norm; that is, $\sup_{a \in \mathcal{A}}\left| h(a) \right| \leq 1$. 
Here, $\left| \psi_{P,t} (h) \right|$ attains its supremum when $h(a) = \mathrm{sign}(\bar{\theta}^a_P(t))$, and the supremum is equal to the probability-weighted $\ell_1$ norm of $\bar{\theta}^A_P(t)$,
\begin{align}
\label{eq:opt_l1}
    \left| \psi_{P,t} (h) \right| 
    = \left| E_P \left\{ \bar{\theta}_P^A(t) \; h(A)\right\} \right|
    \leq E_P \left\{ \left| \bar\theta_{P}^A(t) \right| \right\}
    =E_{P}\left\{\bar{\theta}^A_P(t) \mathrm{sign}\left(\bar{\theta}^A_P(t)\right)\right\} 
    \ .
\end{align} 
Therefore, $\Psi_{P,t}(\mathcal{H})$ is the probability-weighted $\ell_1$ norm of $\bar{\theta}^A_{P}(t)$ whenever $\mathcal{H}$ contains $\mathrm{sign}\left(\bar{\theta}^a_P(t)\right)$. 
Similarly, we can also constrain $h$ to have variance bounded above by one; that is, $\mbox{var}\{h(A)\} \leq 1$. By the Cauchy-Schwarz inequality, the supremum of $|\psi_{P, t} (h)|$ is equal to the variance of $\bar{\theta}^A_P(t)$. The supremum is achieved by $h(a) = \mathrm{var}^{-1/2}\{\bar{\theta}^A_P(t)\}\bar{\theta}^a_P(t)$:
\begin{align*}
     \left| \psi_{P,t} (h) \right| 
    = \left| E_P \left\{ \bar{\theta}_P^A(t) \; h(A)\right\} \right|
    \leq \left| E_P \left\{\bar\theta_{P}^A(t)^2 \right\} \right| ^{1/2} 
    \ .
\end{align*} 
Thus, $\Psi_{P,t}(\mathcal{H})$ is the probability-weighted $\ell_2$ norm of $\bar{\theta}_P^a(t)$ whenever $\mathcal{H}$ contains $\mathrm{var}^{-1/2}\{\bar{\theta}^A_P(t)\}\bar{\theta}^a_P(t)$. 
It is therefore possible to obtain well-behaved estimators for $\ell_1$ or $\ell_2$ norms of $\bar{\theta}_P^a(t)$ in the continuous exposure setting, which overcome the difficulties described in Section \ref{sec:overview}.

Second, we impose a \emph{structural constraint} on $\mathcal{H}$ to restrict its complexity.
Here, we can leverage structural knowledge about $\bar{\theta}_0^a(t)$ to identify constraints on $\mathcal{H}$ that can lead to well-powered tests.
In what follows, we provide some examples of structural constraints that may be used when $\mathcal{H}$ has either the supremum constraint or variance constraint described above.


First, suppose $\mathcal{H}$ has a supremum norm constraint. It is often warranted to assume that $\bar{\theta}^a_{P}(t)$ is monotone, meaning that risk is either non-increasing or non-decreasing as the exposure level increases. 
In this case, if $\bar{\theta}^a_P(t)$ is non-zero, there will exist an exposure level $a_0$ at which point $\bar{\theta}^a_P(t)$ changes sign.
That is, if $\theta^a_P(t)$ is non-decreasing, $\sup_{a \leq a_0} \bar{\theta}^a_P(a) \leq 0$ and $\inf_{a \geq a_0} \bar{\theta}^a_P(a) \geq 0$; if $\theta^a_P(t)$ is non-increasing, $\sup_{a \leq a_0} \bar{\theta}^a_P(a) \geq 0$ and $\inf_{a \geq a_0} \bar{\theta}^a_P(a) \leq 0$.
Therefore, $\mathrm{sign}(\bar{\theta}^a_P(t))$ is given by $(-1)^{\mathds{1}(a \leq a_0)}$ or $(-1)^{\mathds{1}(a \geq a_0)}$ for some $a_0$ and is guaranteed to be contained within the class
\begin{align}
    \mathcal{H}:=\left\{h(A)=(-1)^{\ind(A \leq a)}: a \in \mathcal{A} \right\} \cup \left\{h(A)=(-1)^{\ind(A\geq a)}: a \in \mathcal{A} \right\} .\label{eq:indicators}
\end{align}
From \eqref{eq:opt_l1}, $\Psi_{P,t}(\mathcal{H})$ is equal to the probability-weighted $\ell_1$ norm when $\bar{\theta}^a_P(t)$ is monotone.
Moreover, $\mathcal{H}$ is a class with bounded variation and is known to be a Donsker class \citep[see Example 19.11 of][]{van2000asymptotic}, so efficient estimation of $\psi_{P,t}(h)$ over $\mathcal{H}$ is possible.
In general, when $\bar{\theta}^a_P(t)$ changes sign $K$ times, $\mathrm{sign}(\bar{\theta}^a_P(t))$ is contained in a class of functions with variation norm bounded above by $2K$.
Therefore, we can set 
\begin{align}
\mathcal{H} := \left\{h: \int |dh(a)| \leq 2K, \sup_{a \in \mathcal{A}} |h(a)| \leq 1 \right\} \ .
\label{eq:tvnorm-l1}
\end{align}

Suppose now that $\mathcal{H}$ has a variance constraint.
In this case, in order for $\Psi_{P,t}(\mathcal{H})$ to be equivalent to an $\ell_2$ norm, we need $\mathcal{H}$ to contain a function with unit variance that is proportional to $\bar{\theta}^a_{P}(t)$.
If $\bar{\theta}^a_P(t)$ is monotone, we can set $
\mathcal{H} := \{h: h \text{ is monotone}, \mathrm{var}(h) \leq 1\}$.
More generally, $\mathcal{H}$ can be selected as a class with variance bounded above by one and variation norm bounded above by a constant $\lambda$.
However, unlike the previous case in which $\mathcal{H}$ has a supremum norm constraint, it is difficult to connect a specified bound on the variation norm with an interpretable structural assumption on $\bar{\theta}^a_{P}(t)$.

We reiterate that uniformly Gaussian estimation of $\psi_{P,t}(h)$ ensures nominal type-1 error control for any choice of $\mathcal{H}$ above, and we require no structural knowledge about $\bar{\theta}^a_P(t)$. Moreover, our test is well-powered against alternative hypotheses for which $\Psi_{P,t}(\mathcal{H})$ can be interpreted as a norm.
We thus recommend selecting $\mathcal{H}$ as the smallest function class for which $\Psi_{P,t}(\mathcal{H})$ is nearly equal to a norm for all candidate $\bar{\theta}^a_{P}(t)$ in a range of plausible alternative hypotheses.

\section{Implementation} \label{sec:implementation}

\subsection{Nuisance parameter estimation} \label{sec:nuisance}
Calculating the one-step estimator described in Section \ref{sec:estimation} requires estimation of the following nuisance parameters: the conditional probability of survival, the conditional probability of censoring, and the conditional density of the exposure.
There are many methods for estimating the conditional survival and censoring probabilities. One widely-used semi-parametric regression method is to use the Cox proportional hazard model in conjunction with the Breslow estimator for the baseline cumulative hazard \citep{cox1972regression}. More flexible methods, such as additive Cox regression and survival random forests \citep{hastie2017generalized,ishwaran2008random} can also be used, and multiple methods can be aggregated using an ensemble learning algorithm such as the survival super learner \citep{westling2023inference}.
The conditional density function can be estimated using a kernel smoothing method \citep{ hudson2023approach}. 
This method approximates the conditional density as the conditional expectation of a kernel function and estimates the conditional expectation via the highly adaptive lasso, 
a nonparametric regression estimator \citep{benkeser2016highly}.

\subsection{\texorpdfstring{Calculation of supremum statistic for candidate $\mathcal{H}$}{Calculation of supremum statistic for candidate H}}
\label{sec:implement-class}


We calculate the supremum statistic $\Psi_{n,t}$ using  numerical approximations for the function classes described in Section \ref{sec:construction-of-H}.
For a large integer $\kappa$, let $a_0 < a_1, \ldots, a_\kappa$,  where $a_0$ and $a_\kappa$ are the lower and upper bounds of $\mathcal{A}$, respectively.
To approximate the class in \eqref{eq:indicators}, which allows for $\Psi_{P,t}$ to be interpreted as an $\ell_1$ norm when $\bar{\theta}^a_{P}(t)$ is monotone, we use
\begin{align}
 \tilde{\mathcal{H}}:=\left\{h(A)=(-1)^{\ind(A \leq a_j)}: j = 1, \ldots, \kappa \right\} \cup \left\{h(A)=(-1)^{\ind(A\geq a_j)}: j = 1, \ldots, \kappa \right\} .\label{eq:implement-indicators}
 \end{align}
In this case, $\Psi_{n,t}(\tilde{\mathcal{H}})$ can be easily calculated as the maximum of a finite set of values.

The other specifications of $\mathcal{H}$ described in Section \ref{sec:construction-of-H} can be approximated using a basis expansion.
Let $b_j(a) = \ind\{ a \in [a_{j-1}, a_j)\}$ for $j={1,\dots, \kappa}$.
We consider classes containing functions of the form $h = \sum_{j=1}^{\kappa} \beta_j b_j$, where $\beta = (\beta_1, \dots, \beta_\kappa)^\top \in \mathbb{R}^\kappa$ is a vector of coefficients, and constraints are placed on $\beta$ to enforce scale and structural constraints on $h$.
In particular, one can enforce the supremum constraint $\sup_{a\in\mathcal{A}} |h(a)| \leq 1$ by requiring that $|\beta_j| \leq 1$ for all $j$, or enforce the variance constraint $\mathrm{var}(h) \leq 1$ by requiring that the empirical variance of $h$ is bounded above by one.
To enforce monotonicity, we impose the constraint $\beta_1 \leq \beta_2 \leq \cdots \beta_\kappa$, and to enforce bounded variation, we require $\sum_{j = 2}^\kappa \left|\beta_{j} - \beta_{j-1}\right| \leq \lambda$.
As an example, under supremum and variation constraints, we would obtain the class
\begin{align*}
    \tilde{\mathcal{H}} := \left\{ h = \sum_{j = 1}^\kappa \beta_j b_j: \max_{j = 1, \ldots, \kappa} |\beta_j| \leq 1, \sum_{j=2}^\kappa |\beta_{j} - \beta_{j-1}|\leq \lambda \right\}.
\end{align*}
Our proposed one-step estimator is linear in $h$, so for any $h = \sum_j \beta_j b_j$, we have $\psi^\dagger_{n, t} (h) = \sum_{j} \beta_j \psi^\dagger_{n, t} (b_j)$. 
This allows for the supremum statistic $\Psi^\dagger_{n,t}(\tilde{\mathcal{H}})$ to be characterized as the optimum in a convex optimization problem,
\begin{align*}
    \Psi^\dagger_{n,t}(\tilde{\mathcal{H}}) = \max\left\{\sup_{h \in \tilde{\mathcal{H}}}\psi_n^\dagger(h), -\inf_{h \in \tilde{\mathcal{H}}}\psi_n^\dagger(h) \right\}.
\end{align*}
This problem can be solved using publicly available tools such as the CVXR package in R \citep{fu2020cvxr}. 

We conclude with a comment on tuning parameter selection.
In practice, the complexity of the class $\tilde{\mathcal{H}}$ described above can be tuned by either adjusting the stringency of the structural penalty (e.g., changing the bound $\lambda$ on the variation norm) or by changing the dimension of basis functions (i.e., changing $\kappa$).
Our method satisfies the requisite Donsker assumption on $\mathcal{H}$ so long as one of these tuning parameters is bounded. That is, $\kappa$ can be arbitrarily large if a suitable structural constraint is imposed, and no structural constraint is needed if $\kappa$ is fixed.
The impact of tuning parameter selection on the performance of our test was assessed in simulations.

\subsection{Approximation of null limiting distribution}

We approximate the null limiting distribution of $n^{1/2}\Psi_{n,t}^\dagger(\mathcal{H})$ using Monte Carlo sampling.
For $u = 1, \ldots, U$ and $U$ large, we generate independent draws $m^{(u)}$ from a mean zero Gaussian process with covariance $\Sigma_n=\{\frac{1}{n} \sum_i \bar{D}_n(O_i; h_1)\bar{D}_n(O_i; h_2): {h_1, h_2 \in \mathcal{H}}\}$, where $\bar{D}_n(O_i; h) = 
D_n(O_i;h) - \frac{1}{n}\sum_{i=1}^n D_n(O_i; h)$, and $D_n$ is the empirical influence function in \eqref{eq:empirical-eif}.
Then $M^{(u)} = \sup_{h \in \mathcal{H}} \left| m^{(u)}(h) \right|$ serves as an approximate draw from the limiting distribution of $n^{1/2}\Psi_{n,t}^\dagger(\mathcal{H})$, and a p-value can be approximated as $\frac{1}{U}\sum_{u = 1}^U \mathds{1}\left(\Psi^{\dagger}_{n,t}(\mathcal{H}) > M^{(u)}\right)$.
In the supplementary materials, we provide additional implementation details for generating $m^{(u)}$ and calculating $M^{(u)}$ when $\mathcal{H}$ is selected as one of the function classes in Section 5.2.

\section{Simulation study}
\label{sec:simulation}
We conducted a simulation study to examine finite-sample performance of the proposed method in terms of nominal type-1 error control and statistical power. 
We start by describing our simulation design.

We generated the covariates $W = (W_1, W_2) \in \mathbb{R}^2$ as follows: $W_1$ is from a uniform distribution on $(1, 2)$, and $W_2$ is from a Bernoulli distribution with an equal probability of 0.5 for each outcome. We used different procedures for generating the continuous exposure $A$ depending on whether the null hypothesis held. Under the null, we drew $A$ from a uniform distribution on $(-1, 1)$ independently of $W$.
Under any alternative hypothesis, we set the conditional density of $A$ given $W$ as
\[
g(a \mid w) = \frac{\mbox{expit}\left(\alpha(w)a\right)}{\int_{-1}^1 \mbox{expit}\left(\alpha(w) x \right) dx} \alpha(w) \ind(-1 \leq a \leq 1) \ ,
\]
where $\alpha(w) = 5\left\{ \mbox{expit}(-1 + w_1 - w_2) -0.5 \right\}$, and we generated $A$ using the inverse cumulative distribution function method.

Next, we specified the conditional distributions of survival and censoring, given the covariates, as exponential random variables.
The relationship between the survival/censoring time and the exposures and covariates was governed by two functions $f_1(W)$ and $f_2(A)$.
We fixed $f_1(W) = -3 + 0.3 W_1 + 1.1W_2$, and changed $f_2(A)$ across simulation settings.
Survival and censoring times were then generated under three settings:
\begin{enumerate}[(A)]
    \item \textit{Null}: $T \mid A,W \sim 10\,\mbox{exp}{\left\{0.2f_1(W)\right\}}$ and $C \mid A,W \sim 9\,\mbox{exp}{\left\{-0.2+0.4f_1(W)\right\}}$;
    \item \textit{Alternative \#1}: $T \mid A,W \sim 3.5\,\mbox{exp}{\left\{0.6f_1(W)-0.75f_2(A)\right\}}$ and $C \mid A,W \sim 3.15\,\mbox{exp}{\left\{-1.2+0.4f_1(W)-0.5f_2(A)\right\}}$, with $f_2(A) = A$\sloppy;
    \item \textit{Alternative \#2}: $T \mid A,W \sim 3.5\,\mbox{exp}{\left\{0.6f_1(W)-0.75f_2(A)\right\}}$ and $C \mid A,W \sim 3.15\,\mbox{exp}{\left\{-1.2+0.4f_1(W)-0.5f_2(A)\right\}}$, with $f_2(A)=1.2 - 2 A^2$ \sloppy
\end{enumerate}
In setting (A), the conditional distribution of $T$ did not depend on $A$, so the flat null held.
In setting (B), the alternative held, and the conditional survival probability had a monotone increasing relationship with the exposure.
In setting (C), the alternative again held, and there was a non-monotone quadratic relationship between the exposure and survival probability.
For all simulation settings, we rounded survival and censoring time up to the nearest integer and truncated the censoring time $C$ at $\tau = 35$. We investigated the exposure effects at the specific time point $t=25$.

We estimated the conditional survival and censoring functions using the survival super learner with the following candidate estimators: a Kaplan-Meier estimator, a Cox proportional-hazards regression estimator, a generalized additive Cox proportional-hazards regression estimator, and survival random forests.
We estimated the conditional density of the exposure using the kernel estimator described in Section \ref{sec:nuisance}.
We performed our proposed tests with several specifications of $\mathcal{H}$.
First, we used the specification in \eqref{eq:implement-indicators} and considered $\kappa \in \{10, 20, 50, n\}$.
We also set $\mathcal{H}$ as a class of monotone functions with empirical variance bounded above by one, where the class of monotone functions was approximated using a basis expansion with $\kappa \in \{10, 20, 50\}$ knots.
Additionally, we specified $\mathcal{H}$ as a class with bounded variation norm, under either a supremum norm or variance constraint, as discussed in Section \ref{sec:implement-class}; in each case we used $\kappa \in \{10, 20, 50\}$ basis functions and considered variation bound $\lambda \in \{4, 6\}$.
Finally, we compared with an approach that places no constraint on the variation norm (i.e., $\lambda = \infty$).
All tests were performed at the 0.05 level.

\begin{figure}
    \centering
    \includegraphics[width=\linewidth]{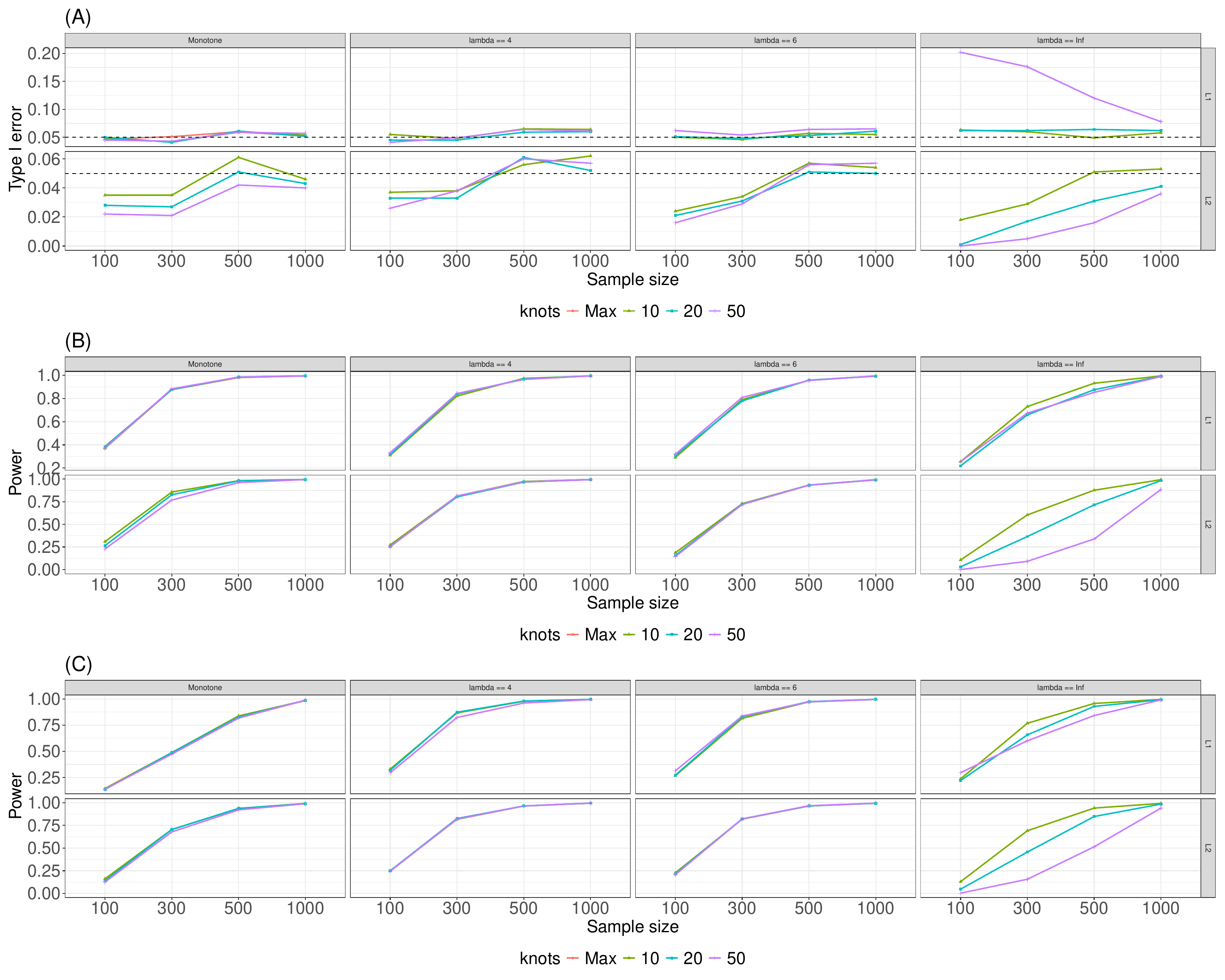}
    \caption{Type-1 error and statistical power for three proposed data generating mechanisms: (A) under the null, (B) under the alternative hypothesis -- monotone effect and (C) under the alternative hypothesis -- non-monotone quadratic effect. The dashed line in (A) corresponds to the nominal level of 0.05.}
    \label{fig:simulation}
\end{figure}

We analyzed 1000 synthetic datasets under each setting above for $n \in \{ 100, 300, 500, 1000\}$. Results are summarized in Figure \ref{fig:simulation}. 
In setting (A), where the null held, all methods under consideration achieved nominal type-1 error control for $n$ large enough. However, when $\mathcal{H}$ was selected as an unstructured class (i.e., $\lambda = \infty$) with a supremum norm scale constraint, and a large number of basis functions was used (i.e., $\kappa = 50$), we observed drastic type-1 error inflation in small samples. In setting (B), as the sample size increased, we achieved good statistical power as long as a structural constraint on $\mathcal{H}$ was used or the number of basis functions was small.
Generally, using structural constraints on $\mathcal{H}$ improved power, compared with the unstructured tests.
In setting (C), we observed that placing variation norm constraints on $\mathcal{H}$ resulted in the most powerful tests.
This was expected because using a more flexible function class should be advantageous when an effect is non-monotone.

\section{Application to HIV vaccine trial}
\label{sec:application}
The proposed method was applied to the pair of Antibody Mediated Prevention (AMP) trials that evaluated prevention efficacy of the HIV monoclonal antibody VRC01: HIV Vaccine Trials Network (HVTN) 704/HIV Prevention Trials Network (HPTN) 085 and HVTN 703/HPTN 081 
conducted in the Americas and in southern Africa, respectively \citep{corey2021two,seaton2023pharmacokinetic}. 
The primary objective compared the survival probability of new HIV-1 diagnosis by 80 weeks between the VRC01 arms pooled vs. placebo \citep{corey2021two}.  A secondary objective was to assess, among VRC01 recipients, the effect of VRC01 serum concentration at Day 61 on the time from Day 61 to new HIV-1 diagnosis by 80 weeks, operationalized as Day 500 post Day 61. Participants were sampled for measurement of VRC01 concentration using a two-phase case-control sampling design with known sampling weights. All participants in a VRC01 arm who acquired the HIV-1 diagnosis endpoint ($N=85$ cases) were included in the analysis subset and assigned a weight of one. In addition, 80 non-case participants were sampled for concentration measurement. We thereby included 165 VRC01 recipients (82 from HVTN 703 and 83 from HVTN 704) with Day 61 VRC01 concentration data, who also had VRC01 concentrations data measured every 4 weeks through 80 weeks. 
The goal of the analysis was to determine whether the counterfactual probability of survival by Day 500 post Day 61 varied across different levels of Day 61 VRC01 concentration, pooling data across the two trials. 
We adjusted for the following two baseline covariates that may either confound the exposure-outcome relationship or improve prediction efficiency: body weight and geographic region (South Africa, Southern Africa outside of South Africa, Peru or Brazil, U.S. or Switzerland).

\begin{figure}
    \centering
    \includegraphics[width=\linewidth]{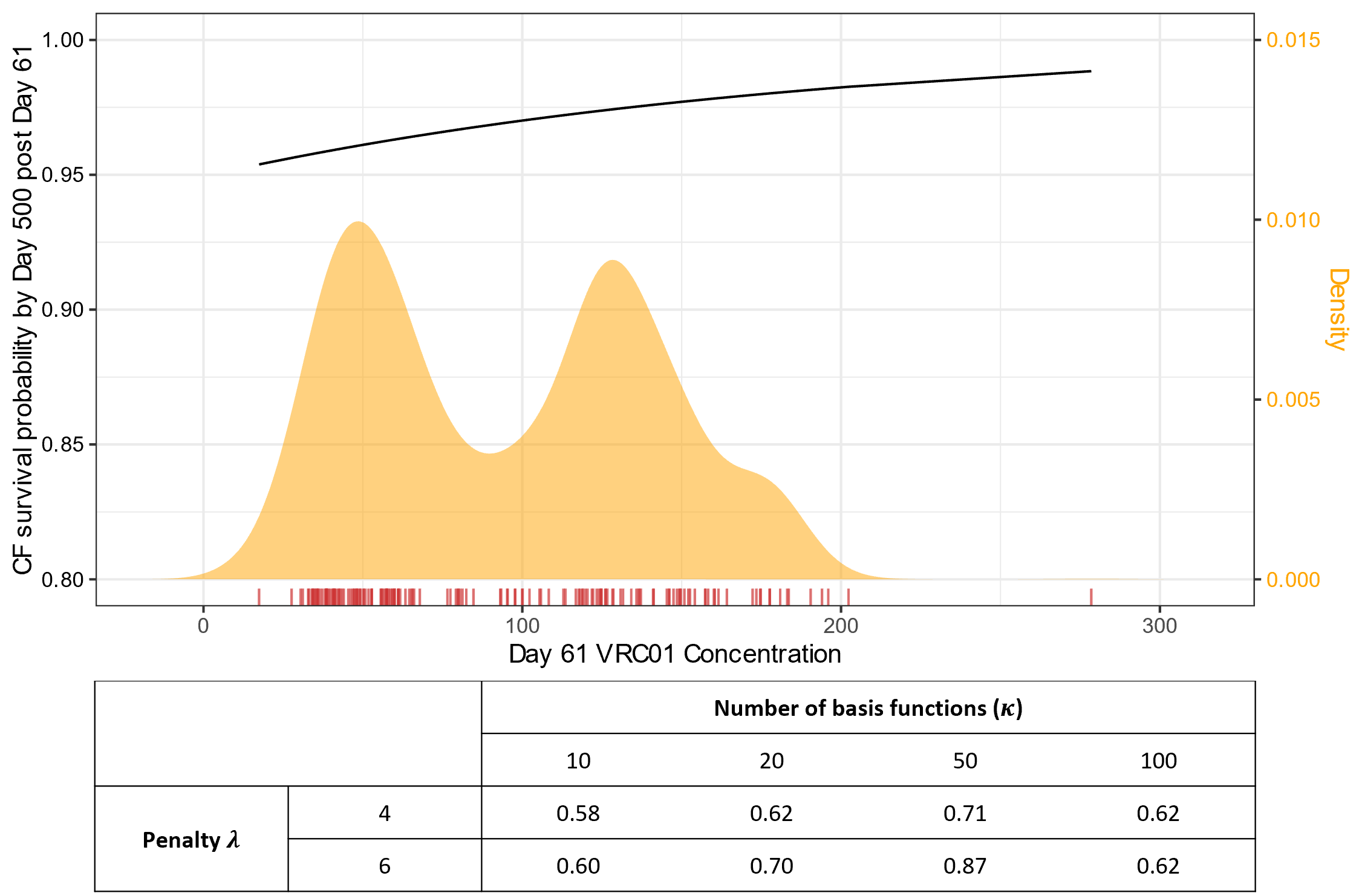}
    \caption{The black curve shows a plug-in estimate of the counterfactual probability of survival at Day 500 post Day 61 across levels of Day 61 VRC01 concentration. The gold curve shows the distribution of Day 61 VRC01 concentration.
    Both the plug-in survival probability and density estimates were obtained via inverse probability weighting, with respect to the probability of being selected to have Day 61 VRC01 concentration measured.
    The table shows p-values resulting from our proposed test with different tuning parameters.}
    \label{fig:data_application}
\end{figure}

Figure \ref{fig:data_application} presents a plug-in estimate of the counterfactual survival probabilities at Day 500 post Day 61 across varying Day 61 VRC01 concentrations. The counterfactual survival curve was nearly flat, indicating no strong association between survival probabilities and Day 61 VRC01 concentration. To formally assess whether this effect existed across concentrations, we applied our proposed hypothesis test with different combinations of tuning parameters: $\kappa\in\{10, 20, 50, 100\}$ and $\lambda\in\{4, 6\}$. The resulting p-values are also presented in Figure \ref{fig:data_application}. All p-values were non-significant, further suggesting that there was no strong effect. 
These findings suggest that VRC01 concentration measured at Day 61 is not a strong predictor of acquisition of HIV-1. 
Within individuals, the VRC01 concentration had a strong sawtooth pattern across the 10 infusions,
and the hazard of new HIV-1 diagnosis significantly depended on the current value of VRC01 concentration as modeled pharmacokinetically \citep{seaton2023pharmacokinetic}.
In contrast, by focusing on VRC01 concentration at a single time point, the current analysis accesses information about a correlate solely from inter-individual variability, not also accounting for information in the intra-individual sawtooth temporal variability that enabled the time-varying modeling to detect a correlate.  The finding that the single early time point cannot be used to obtain an adequate predictor of HIV-1 acquisition is unfortunate because such a finding would constitute progress toward 
development of a surrogate endpoint; thus this finding is significant in providing the insight that temporal modeling of VRC01 concentration is needed for improving the predictive biomarker.

\section{Discussion}
\label{sec:discussion}
We have proposed a novel nonparametric hypothesis testing framework for evaluating the effect of a continuous exposure on a right-censored survival outcome.
Our test is made robust by leveraging flexible machine learning tools to adjust for confounding factors without strong distributional assumptions.
Our simulation experiments indicate proper type-1 error control even in small sample settings.

We have demonstrated that our test is powerful and can identify complex (e.g., nonlinear) relationships between the exposure and survival time.
However, to attain good power, careful construction of the contrast class $\mathcal{H}$ is needed.
It is currently unclear how to optimally specify $\mathcal{H}$, so we require future research on tuning parameter selection methods. 
In our current paper, we provide examples of function classes that can yield good power in realistic settings.
This heuristic guidance can make our inference procedure practical when pre-specified statistical analyses are required.

\section*{Data availability statement}
The supporting code is available on \href{https://github.com/YutongJ/Nonparametric-methods-for-evaluating-the-effect-of-continuous-treatments-on-survival-outcomes}{GitHub}, and the trial data used in the analysis is publicly accessible at \href{https://atlas.scharp.org/project/HVTN%20Public%20Data/begin.view}{SCHARP Atlas}.

\section*{Acknowledgments}
We thank the AMP study participants and AMP study team. This work was supported by the National Institute of Allergy and Infectious Diseases of the NIH, under award UM1AI068635, by start-up funds from the Fred Hutchinson Cancer Center, and by the Office of Research Infrastructure Programs, NIH (award S10OD028685 to Fred Hutchinson Cancer Center).  The content is solely the responsibility of the authors and does not necessarily represent the official views of the NIH.






\label{lastpage}



\newpage
\section*{Supplementary Materials}














\subsection*{1. Proofs of Theoretical Results} \label{proofs}

\subsubsection*{Proof of Theorem 1}



Let $P$ be a probability distribution that satisfies the causal identification conditions $(A1)-(A4)$, and let $p$ be the density of $P$. 
We define $\Pe$ as the parametric submodel with $P_{\epsilon=0}=P$ and score function $\omega(\cdot)$.
 To be explicit, we define the density function of $\Pe$ as
\[\pe(o) = p(o)[1 + \epsilon\omega(o)].\]
Any distribution in a neighborhood of $P$ can be approximated by such a submodel.
We let denote $Q_\epsilon(W)$ the marginal distribution of $W$ under$\Pe$.

To argue that $\psi_{P,t}(h)$ is pathwise differentiable, it suffices to show that for any score $\omega$,
\[\left. \frac{\partial}{\partial \epsilon} \psi_{\Pe}(h) \right|_{\epsilon = 0}  = E[D_P(O; h)\omega(O)] \ ,
\]
where $D_{P}(O;h)$ is centered to have mean zero.

First, we write $\psi_{\Pe, t}(h)$ as a sum of two components:
\begin{align*}
    \psi_{\Pe, t}(h) 
    = E_{\Pe}\left\{ \bar\theta^A(t)_{P_\epsilon} h(A) \right\}
    = G_{\Pe, t}^{\mathrm{I}}(h)
    - G_{\Pe, t}^{\mathrm{II}}(h),
\end{align*}
where we let
\begin{align*}
    G_{\Pe, t}^{\mathrm{I}}(h)  &= E_{\Pe}\left\{ \theta_{P_\epsilon}^A(t) h(A) \right\} \ , \\
    G_{\Pe, t}^{\mathrm{II}}(h) &= E_{\Pe}\left\{ \theta_{P_\epsilon}^A(t)\right\} E_{\Pe}\left\{ h(A) \right\}.
\end{align*}

We first evaluate the derivative of the first additive component $G_{\Pe, t}^{\mathrm{I}}(h)$ at $\epsilon=0$.
We can re-write $G_{\Pe, t}^{\mathrm{I}}(h)$ as
\begin{align}
    G_{\Pe, t}^{\mathrm{I}}(h)  
    =& \int \theta_{\Pe}^a(t) h(a) p(w,a,y,\delta)[1 + \epsilon \omega(w,a,y,\delta)] dP(w,a,y,\delta) \ .
\end{align}
Evaluating the derivative of $G_{\Pe, t}^{\mathrm{I}}(h)$ requires calculating the derivative of $\theta^a_{P_\epsilon}(t)$. We have
\begin{align}
    \left.\frac{\partial}{\partial\epsilon} \theta_{\Pe}^a(t) \right|_{\epsilon=0}
    =& \left.\frac{\partial}{\partial\epsilon} E_{\Pe}[S_P(t \mid A=a, W)] \right|_{\epsilon=0} \nonumber\\
    =& \left.\frac{\partial}{\partial\epsilon} \int S_{P_\epsilon}(t\mid a,w) p_\epsilon(w) P(dw) \right|_{\epsilon=0} \nonumber\\
    =& \int \left. \frac{\partial}{\partial\epsilon} S_{P_\epsilon}(t\mid a,w) \right|_{\epsilon=0} dQ(w) + \int S_P(t\mid a,w) \left. \frac{\partial}{\partial\epsilon} p_\epsilon(w) \right|_{\epsilon=0} P(dw) \ . \nonumber
\end{align}
Using the chain rule and the fact that $S_{P_\epsilon}(t \mid a,w) := \prodi_{(0,t]} \{ 1- \Lambda_{P_\epsilon}(du \mid a,w)\}$,we have
\begin{align}
    \int \left. \frac{\partial}{\partial\epsilon} S_{P_\epsilon}(t\mid a,w) \right|_{\epsilon=0} dQ(w)
    =& \int \left. \frac{\partial}{\partial\epsilon} \prodi_{(0,t]}  \left[ 1-\Lambda_{P_\epsilon} (du \mid a,w) \right] \right|_{\epsilon=0} dQ(w) \nonumber\\
    =& \int -S_P(t\mid a,w) \int_0^t \frac{S_P(u^-\mid a,w)}{S_P(u \mid a,w)} \left. \frac{\partial}{\partial\epsilon} \Lambda_{P_\epsilon} (du \mid a,w) \right|_{\epsilon=0} dQ(w) \ . \label{firstterm}
\end{align}
We now define 
\begin{align*}
    F_{\epsilon,1}(t \mid a,w) &= P_\epsilon(Y\leq t, \Delta=1 \mid a,w)\\
    R_\epsilon(t\mid a,w) &= P_\epsilon(Y \geq t \mid a,w) \ .
\end{align*}
We also observe that
\begin{align}
    \left. \frac{\partial}{\partial\epsilon} p_\epsilon(y,\delta \mid a,w) \right|_{\epsilon=0} &= p(y,\delta \mid a,w) \omega(y,\delta \mid a,w) \label{dens1}\\
    \left. \frac{\partial}{\partial\epsilon} p_\epsilon(w) \right|_{\epsilon=0} &= p(w) \omega(w) \label{dens2} \ .
\end{align}
Since we have
\begin{align*}
    \left. \frac{\partial}{\partial\epsilon} \Lambda_{P_\epsilon} (t \mid a,w) \right|_{\epsilon=0}
    =& \left. \frac{\partial}{\partial\epsilon}\int_0^t \frac{F_{\epsilon,1}(du \mid a,w)}{R_\epsilon(u\mid a,w)} \right|_{\epsilon=0}\\
    =& \int_0^t R(u\mid a,w)^{-1} \left. \frac{\partial}{\partial\epsilon} F_{\epsilon,1}(du \mid a,w) \right|_{\epsilon=0}\\
    &- \int_0^t \left. \frac{\partial}{\partial\epsilon} R_\epsilon (u \mid a,w) \right|_{\epsilon=0} R(u\mid a,w)^{-2} F_{0,1} (du\mid a,w) \ ,
\end{align*}
we can write
\begin{align*}
    \left. \frac{\partial}{\partial\epsilon} \Lambda_{P_\epsilon} (du \mid a,w) \right|_{\epsilon=0}
    = \frac{\left. \frac{\partial}{\partial\epsilon} F_{\epsilon,1}(du\mid a,w) \right|_{\epsilon=0}}{R(u \mid a,w)} - \frac{\left. \frac{\partial}{\partial\epsilon} R_\epsilon(u\mid a,w) \right|_{\epsilon=0} F_{0,1}(du \mid a,w)}{R(u\mid a,w)^2} \ .
\end{align*}
In addition, 
\begin{align*}
    \left. \frac{\partial}{\partial\epsilon} F_{\epsilon,1}(du \mid a,w) \right|_{\epsilon=0}
    =& \left. \frac{\partial}{\partial\epsilon} P_\epsilon(Y \leq u, \Delta=1 \mid a,w) \right|_{\epsilon=0} \\
    =& \left. \frac{\partial}{\partial\epsilon} \iint \ind(y\leq u, \delta=1) P_\epsilon(dy, d\delta \mid a,w) \right|_{\epsilon=0} \\
    =& \iint \ind(y\leq u, \delta=1) \left. \frac{\partial}{\partial\epsilon} P_\epsilon(dy, d\delta \mid a,w) \right|_{\epsilon=0} \\
    =& \iint \ind(y\leq u, \delta=1) \omega(y,\delta \mid a,w) P(du,d\delta\mid a,w)\\
    \left. \frac{\partial}{\partial\epsilon} R_{\epsilon}(u \mid a,w) \right|_{\epsilon=0}
    =& \left. \frac{\partial}{\partial\epsilon} P_{\epsilon}(Y \geq u \mid a,w) \right|_{\epsilon=0}\\
    =& \left. \frac{\partial}{\partial\epsilon} \iint \ind(Y \geq u) P_\epsilon( dy, d\delta \mid a,w) \right|_{\epsilon=0} \\
    =& \iint \ind(Y \geq u) \omega(y,\delta \mid a,w) P( dy, d\delta \mid a,w) \ .
\end{align*}
We can further write (\ref{firstterm}) as follows:
\begin{align*}
    & \int \left. \frac{\partial}{\partial\epsilon} S_{P_\epsilon}(t\mid a,w) \right|_{\epsilon=0} dQ(w)\\
    =& \int -S_P(t\mid a,w) \int_0^t \frac{S_P(u^-\mid a,w)}{S_P(u \mid a,w)} \frac{\left. \frac{\partial}{\partial\epsilon} F_{\epsilon,1}(du\mid a,w) \right|_{\epsilon=0}}{R(u \mid a,w)} dQ(w)\\
    &+ \int -S_P(t\mid a,w) \int_0^t \frac{S_P(u^-\mid a,w)}{S_P(u \mid a,w)} \frac{\left. \frac{\partial}{\partial\epsilon} R_\epsilon(u\mid a,w) \right|_{\epsilon=0} F_{0,1}(du \mid a,w)}{R(u\mid a,w)^2} dQ(w)\\
    =& - \iiint \ind(u\leq t, \delta=1) \frac{S_P(t\mid a,w)S_P(u^-\mid a,w)}{S_P(u\mid a,w)R(u\mid a,w)} \omega(u,\delta\mid a,w)P(du, d\delta\mid a,w)dQ(w)\\
    &+ \iiiint \ind(y\geq u, u \leq t) \frac{S_P(t\mid a,w)S_P(u^-\mid a,w)}{S_P(u\mid a,w)R(u\mid a,w)} \omega(u,\delta\mid a,w)P(du, d\delta\mid a,w) F_{0,1}(du \mid a,w) dQ(w) \ .
\end{align*}

Using the fact that
\begin{align*}
    \int S_P(t\mid a,w) \left. \frac{\partial}{\partial\epsilon} p_\epsilon(w) \right|_{\epsilon=0} P(dw)
    =& \int S_P(t\mid a,w) \omega(w) p(w) P(dw) = \int S_P(t\mid a,w) \omega(w) dQ(w) \ , 
\end{align*}
we can express the derivative of $G_{\Pe, t}^{\mathrm{I}}(h)$ as
\begin{align}
    & \left.\frac{\partial}{\partial\epsilon} G_{\Pe, t}^{\mathrm{I}}(h) \right|_{\epsilon=0} 
 \nonumber\\
    =& \int \left. \frac{\partial}{\partial\epsilon} \theta_{\Pe}^a(t) \right|_{\epsilon=0} h(a) dP(w,a,y,\delta) + \int  \theta_{P}^a(t) h(a) \omega(w,a,y,\delta) dP(w,a,y,\delta) \nonumber\\
    =& -\iiiint \ind(u\leq t, \delta=1) \frac{S_P(t| a,w)S_P(u^-| a,w)}{S_P(u| a,w)R(u| a,w)} \omega(u,\delta| a,w)P(du, d\delta| a,w)dQ(w) h(a) dP(w,a,y,\delta) \nonumber \\
    &+ \int\iiiint \ind(y\geq u, u \leq t) \frac{S_P(t| a,w)S_P(u^-| a,w)}{S_P(u| a,w)R(u| a,w)} \omega(u,\delta| a,w)P(du, d\delta| a,w)\times \nonumber \\
    & F_{0,1}(du | a,w) dQ(w) h(a) dP(w,a,y,\delta) \nonumber \\
    =& E\left\{ E\left[ S_P(t|A,W) \frac{p(A)}{p(A|W)} E\left\{ H(t \wedge Y, A,W)\omega(Y,\Delta,A,W)|A,W\right\}|A \right]h(A)\right\} \nonumber \\
    &- E\left\{ E\left[ S_P(t|A,W) \frac{p(A)}{p(A|W)} E\left\{ \frac{\ind(Y \leq t, \Delta=1) S_P(Y^-| A,W)}{S_P(Y| A,W) R(Y|A,W)}\omega(Y,\Delta,A,W)|A,W\right\}|A \right]h(A)\right\} \nonumber \\
    &+ E\left\{ E\left[ S_P(t|A,W) h(A)\right] \omega(Y,\Delta,A,W)\right\} 
    + E\left\{ E\left[ S_P(t|A,W) \frac{p(A)}{p(A|W)}\mid A \right] h(A) \omega(Y,\Delta,A,W) \right\} \ . \label{G1}
\end{align}

By performing a similar calculation, one can show that the derivative of $G_{\Pe, t}^{\mathrm{II}}(h)$ can be expressed as
\begin{align}
    & \left.\frac{\partial}{\partial\epsilon} G_{\Pe, t}^{\mathrm{II}}(h) \right|_{\epsilon=0} 
 \nonumber\\
    =& \left. \frac{\partial}{\partial\epsilon} E_{\Pe}\left\{ \theta^A(t)\right\} E_{\Pe}\left\{ h(A) \right\} \right|_{\epsilon=0} \nonumber \\
    =& \left\{ \int \left. \frac{\partial}{\partial\epsilon} \theta^a_{\Pe}(t) \right|_{\epsilon=0} dP(w,a,y,\delta) \right\} \left\{ h(a) dP(w,a,y,\delta) \right\}\nonumber \\
    &+ \left\{ \int \theta_{P}^a(t) \omega(w,a,y,\delta) dP(w,a,y,\delta) \right\} \left\{ h(a) dP(w,a,y,\delta) \right\}\nonumber \\
    &+ \left\{ \int \theta_{P}^a(t) dP(w,a,y,\delta) \right\} \left\{ h(a) \omega(w,a,y,\delta) dP(w,a,y,\delta) \right\}\nonumber \\
    =& E\left\{ E\left[ S_P(t|A,W) \frac{p(A)}{p(A|W)} E\left\{ H(t \wedge Y, A,W)\omega(Y,\Delta,A,W)|A,W\right\}|A \right]\right\} E\{h(A)\}\nonumber \\
    &- E\left\{ E\left[ S_P(t|A,W) \frac{p(A)}{p(A|W)} E\left\{ \frac{\ind(Y \leq t, \Delta=1) S_P(Y^-| A,W)}{S_P(Y| A,W) R(Y|A,W)}\omega(Y,\Delta,A,W)|A,W\right\}|A \right]\right\} E\{h(A)\}\nonumber \\
    &+ E\left\{ E\left[ S_P(t|A,W)\right] \omega(Y,\Delta,A,W)\right\} E\{h(A)\}\nonumber \\
    &+ E\left\{ E\left[ S_P(t|A,W) \frac{p(A)}{p(A|W)}\mid A \right] \omega(Y,\Delta,A,W) \right\} E\{h(A)\} \nonumber \\
    &+ E\left\{ E\left[ S_P(t|A,W) \frac{p(A)}{p(A|W)}\mid A \right] \right\} E\{h(A)\omega(Y,\Delta,A,W)\} \ . \label{G2}
\end{align}

Finally, by (\ref{G1}) and (\ref{G2}), we can express $\left. \frac{\partial}{\partial\epsilon} \psi_{\Pe, t}(h) \right|_{\epsilon=0}$ as
\begin{align*}
    & \left. \frac{\partial}{\partial\epsilon} \psi_{\Pe, t}(h) \right|_{\epsilon=0} \\
    =& \left.\frac{\partial}{\partial\epsilon} G_{\Pe, t}^{\mathrm{I}}(h) \right|_{\epsilon=0} - \left.\frac{\partial}{\partial\epsilon} G_{\Pe, t}^{\mathrm{II}}(h) \right|_{\epsilon=0} \\
    =& E\left\{ D_P(W,A,Y,\Delta; h) \omega(W,A,Y,\Delta) \right\} \ ,
\end{align*}

where
\begin{align*}
    D_P(o; h) 
    =& \left\{h(a) - E_P[h(A)]\right\} \left\{ \theta_{P}^{a}(t) - E\left[\theta_{P}^{A}(t)\right]\right\} \\
    &+ \left\{h(a) - E_P[h(A)]\right\} \left\{ S_{P}(t \mid a,w) \frac{E_P(g_P(a \mid W))}{g_P(a \mid w)} \left[ H(t \wedge y, a, w) \right] \right\} \\
    &- \left\{h(a) - E_P[h(A)]\right\} \left\{ S_{P}(t \mid a,w) \frac{E_P(g_P(a \mid W))}{g_P(a \mid w)} \left[ \frac{\ind(y \leq t, \delta=1) S_{P}(y^- \mid a,w)}{S_{P}(y \mid a,w) R_{P}(y \mid a,w)} \right] \right\} \\
    &+ E_{P}\left[ S_{P}(t \mid A,w) \{h(A) - E_P[h(A)]\} \right] 
    - E_{P}[h(A)] E[\theta_{P}^A(t)]
\end{align*}

This completes the proof.

\subsubsection*{Proof of Theorem 2}

The one-step estimator can be expressed as
\begin{align*}
    \psi_{n,t}^{\dagger} (h) - \psi_{P,t} (h) = \frac{1}{n} \sum_{i=1}^n D_{P}(O_i; h) + r_n(h) \ .
\end{align*}
We write the remainder term as
\begin{align*}
    r_n(h)
    = r_n^{\mathrm{I}}(h) + r_n^{\mathrm{II}}(h) + r_n^{\mathrm{III}}(h) \ ,
\end{align*}
where we define
\begin{align*}
    r_n^{\mathrm{I}}(h)
    :=& - \int \left\{ \bar\theta_n^a(t) - \bar\theta_P^a(t) \right\} \left\{ h(a) - E[h(A)] \right\} d(P_n - P) \ ,\\
    r_n^{\mathrm{II}}(h)
    :=& \int \left[ \bar\theta_n^a(t) - \bar\theta_P^a(t) \right] \left[ n^{-1}\Sigma_i h(A_i) - E[h(A)] \right] d(P_n - P) \ ,\\
    r_n^{\mathrm{III}}(h)
    :=& \int h(a) \int S_n(t| a,w) \int_0^t \left( \frac{g_P G_P(y| a,w)}{g_n G_n(y^-| a,w)} -1 \right) \left( \frac{S_P(y| a,w)}{S_n(y| a,w)} -1 \right) (dy| a,w) dP_W dP_A\\
    &- \frac{1}{n}\Sigma_i h(A_i) \iint S_n(t|a,w) \int_0^t \left( \frac{g_P G_P(y| a,w)}{g_n G_n(y^-| a,w)} -1 \right) \left( \frac{S_P(y| a,w)}{S_n(y| a,w)} -1 \right) (dy| a,w) dP_W dP_A \,
\end{align*}
where $P_A$ and $P_W$ denote the marginal distributions of $A$ and $W$, respectively.

We first show that $\sup_{h \in \mathcal{H}} \left| r_n^{\mathrm{I}}(h) \right|$ is $o_P(n^{-1/2})$. By Assumptions (B1) and (B2), Theorem 2.10.6 of \citet{van1996weak} and the assumption that $h$ is bounded, the following hold:
\begin{enumerate}
    \item $\sup_{h\in \mathcal{H}} \int \left\{ \bar{\theta}_n^a(t) - \bar{\theta}_P^a(t) \right\}^2 \left\{h(a) - E(h)\right\}^2 = o_P(1)$
    \item $\left\{ \bar\theta_n^a(t) - \bar\theta_P^a(t) \right\} \left\{ h(a) - E[h(A)] \right\}$ falls in a $P$-Donsker class.
\end{enumerate}
It is shown in the proof of lemma 19.26 in \citet{van2000asymptotic} that the first reminder term therefore satisfies
\[ \sup_{h \in \mathcal{H}} \left| r_n^{\mathrm{I}}(h) \right| = \sup_{h\in \mathcal{H}} \int \left\{ \left(\bar\theta_n^a(t) - \bar\theta_P^a(t) \right)\left( h(a) - E[h(A)] \right) \right\} d(P_n - P) = o_P(n^{-1/2}) \ . \]

We now show that $\sup_{h \in \mathcal{H}} \left| r_n^{\mathrm{II}}(h) \right|$ is also $o_P(n^{-1/2})$. 
By Assumption (B1) and Theorem 2.10.6 of \citet{van1996weak}, $\mathcal{H}$ is a $P$-Donsker class, and we have
\[\sup_{h\in \mathcal{H}} \left| \frac{1}{n} \sum_{i=1}^n h(A_i) - E[h(A)] \right| = O_P(n^{-1/2}) \ .\]
From Assumption (B2) and the fact that $h$ is bounded, the following hold:
\begin{enumerate}
    \item  $\sup_{h\in \mathcal{H}} \int \left\{ \bar{\theta}_n^a(t) - \bar{\theta}_P^a(t) \right\}^2 \left\{n^{-1}\Sigma_i h(A_i) - E(h)\right\}^2 = o_P(1) O_P(n^{-1/2}) = o_P(n^{-1/2})$
    \item $\left\{ \bar\theta_n^a(t) - \bar\theta_P^a(t) \right\} \left\{ n^{-1}\Sigma_i h(A_i) - E[h(A)] \right\}$  falls in a $P$-Donsker class.
\end{enumerate}
By again applying \citet{van2000asymptotic} lemma 19.26, we can conclude that $\sup_{h \in \mathcal{H}} \left| r_n^{\mathrm{II}}(h) \right| = o_P(n^{-1/2})$.

Next, we show that $\sup_{h \in \mathcal{H}} \left| r_n^{\mathrm{III}}(h) \right|$ is $o_P(n^{-1/2})$. We can rewrite this final term as
\begin{align*}
    r_n^{\mathrm{III}}(h)
    :=&\iint S_n(t|a,w) \int_0^t \left( \frac{g_P G_P(y| a,w)}{g_n G_n(y^-| a,w)} -1 \right) \left( \frac{S_P(y| a,w)}{S_n(y| a,w)} -1 \right) (dy| a,w) dP_W \\
    & \left[ h(a) - E(h(A))+ E(h(A))- \frac{1}{n}\Sigma_i h(A_i)\right] dP_A\\
    =& \int M(a,t) \left\{ h(a) - E[h(A)]\right\} dP_A\\
    &+ \int M(a,t) \left\{ E[h(A)] - \frac{1}{n}\Sigma_i h(A_i) \right\} dP_A \ ,
\end{align*}
where $M(a,t) = \int S_n(t|a,w) \int_0^t \left( \frac{g_P G_P(y| a,w)}{g_n G_n(y^-| a,w)} -1 \right) \left( \frac{S_P(y| a,w)}{S_n(y| a,w)} -1 \right) (dy| a,w) dP_W$.\\
Using Assumption (B3) and recalling that $h$ is bounded, $\int M(a,t) dP_A$ is $o_P(n^{-1/2})$. Since $h$ is bounded and $\sup_{h\in \mathcal{H}} \left| \frac{1}{n} \sum_{i=1}^n h(A_i) - E[h(A)] \right|$ is $O_P(n^{-1/2})$, we can conclude that $\sup_{h \in \mathcal{H}} \left| r_n^{\mathrm{III}}(h) \right| = o_P(n^{-1/2})$.

This completes the proof that the one-step estimator is uniformly asymptotically linear, i.e., that $\sup_{h \in \mathcal{H}}|r_n(h)| = o_P(n^{-1/2}).$ 
To see that $\{n^{1/2}[\psi_{n,t}^\dagger(h) - \psi_{P,t}(h)]: h \in \mathcal{H}\}$ converges to a Gaussian process, we first observe that Donsker's theorem implies that $\{n^{-1/2}\sum_{i=1}^n D_P(O_i; h): h \in \mathcal{H}\}$ converges weakly to $\{\mathbb{G}(h):h \in \mathcal{H}\}$.  The argument is completed via an application of Slutsky's theorem.

\subsection*{2. Approximation of null limiting distribution of $\Psi_{n,t}^\dagger(\mathcal{H})$}
\label{null_distribution}

In this section, we describe in more detail how we approximate the null limiting distribution of $\Psi_{n,t}^\dagger(\mathcal{H})$ for the choices of $\mathcal{H}$ described in Section 5.2.
We start with the case where 
\begin{align*}
 \tilde{\mathcal{H}} =\left\{h(A)=(-1)^{\ind(A \leq a_j)}: j = 1, \ldots, \kappa \right\} \cup \left\{h(A)=(-1)^{\ind(A\geq a_j)}: j = 1, \ldots, \kappa \right\} .
 \end{align*}
 For $j \in \{1, \ldots, \kappa\}$, let $h_j := (-1)^{\mathds{1}(A \leq a_j)}$, and let $\mathbf{D}_n$ be an $n$ by $\kappa$ matrix with element $(i, j)$ given by $\bar{D}_{n}(O_i; h_j)$, where we recall $\bar{D}_n(o;h) = D_n(o;h) - n^{-1}\sum_{i=1}^n D_n(O_i; h)$. Now, let $V_n = n^{-1} \mathbf{D}_n^\top \mathbf{D}_n$.
 For $u = 1, \ldots, U$ and $U$ larger, let $\boldsymbol{\xi}^{(u)}$ be an $N(0, V_n)$ random vector.
 We approximate a draw from the Gaussian process $\mathbb{G}$ as $m_n^{(u)}$, where 
 \begin{align*}
     m_n^{(u)}(h_j) = \boldsymbol{\xi}_j, \quad m_n^{(u)}(h_j) = -\boldsymbol{\xi}_j
 \end{align*}
 and we take $M_n^{(u)} = \max_{h \in \mathcal{\tilde{H}}} \left|m_n^{(u)}(h)\right|$ as an approximate draw from $\sup_{h \in \tilde{\mathcal{H}}}|\mathbb{G}(h)|$.
  We approximate a p-value as $\frac{1}{U}\sum_{u = 1}^U \mathds{1}\left(\Psi^{\dagger}_{n,t}(\mathcal{H}) > M^{(u)}\right)$.

Suppose now that we use a class that contain functions of the form $h = \sum_j \beta_j b_j$, where  $b_j(a) = \ind\{ a \in [a_{j-1}, a_j)\}$, $\beta = (\beta_1, \beta_2, \ldots, \beta_\kappa) \in \mathbb{R}^\kappa$, and constraints are placed on $\beta$ to ensure that $h$ satisfies specified scale and structural constraints.
Now, let $\mathbf{B}_n$ be an $n$ by $\kappa$ matrix with element $(i,j)$ equal to $\bar{D}_n(O_i; b_j)$.
Let $V_n = n^{-1} \mathbf{B}_n^\top \mathbf{B}_n$, and let $\boldsymbol{\xi}^{(u)}$ be an $N(0, V_n)$ random variable.
For any $h = \sum_j \beta_j b_j$, we set $m_n^{(b)}(h) = \beta^\top \boldsymbol{\xi}^{(b)}$ and then take $M^{(u)} = \sup_{h \in \mathcal{H}} \left|m_n^{(u)}\right|$.
Similarly as in Section \ref{sec:implement-class}, $M^{(u)}$ can be expressed as the optimal value in a convex optimization problem for any of the choices of constraints we consider, making computation fairly simple.



\bibliographystyle{plainnat}
\bibliography{refs}

\end{document}